\documentstyle[aps,epsf,prc]{revtex}

\begin{document}
\title{Associative charm photoproduction with circularly polarized photons}
\author{Michail P. Rekalo \footnote{ Permanent address:
\it National Science Center KFTI, 310108 Kharkov, Ukraine}
}
\address{Middle East Technical University, 
Physics Department, Ankara 06531, Turkey}
\author{Egle Tomasi-Gustafsson}
\address{\it DAPNIA/SPhN, CEA/Saclay, 91191 Gif-sur-Yvette Cedex, 
France}
\date{\today}

\maketitle
\begin{abstract}
We calculate the asymmetries for the processes $\vec\gamma+\vec p\to \Lambda_c^+ (\Sigma_c^+)+\overline{D^0}$ and $\vec\gamma+\vec p\to \Lambda_c^+ (\Sigma_c^+)+\overline{D^*}$, induced by 
circularly polarized photons on a polarized proton target in the framework 
of an effective Lagrangian model. We found large and positive values for the asymmetry induced by the $z-$component of the proton polarization, in a wide energy region, from threshold up to $E_\gamma=70$ GeV. In the case of pseudoscalar meson photoproduction, the predictions are qualitatively model-independent. For $D^*$-photoproduction, there is a strong dependence of the asymmetries on $N\Lambda_c D(D^*)$-coupling constants.
\end{abstract}
\section{Introduction}

The question of how the proton spin is carried by its constituents is still very actual \cite{Ad94a,Ad95,Ab94a,Ab94b}. The determination of the gluon contribution, $\Delta G$, to the nucleon spin is the object of different experiments with polarized beams and targets \cite{Ad94b,Bu92}. In particular, the production of charmed particles in collisions of longitudinally polarized muons with polarized proton target will be investigated by the Compass collaboration \cite{Compass}. In the framework of the photon-gluon fusion model \cite{Ha78}, $\gamma^*+g\to c+\overline{c}$, ($\gamma^*$ is a virtual photon), the corresponding asymmetry (for polarized $\mu p$-collisions), is related to the polarized gluon content in the polarized protons \cite{Gl88}. Due to the future impact of such result, it seems necessary to understand all the other possible mechanisms which can contribute to inclusive charm photoproduction, such as $\gamma+p\to \overline{D^0}+X$, for example. One possible and $a~priori$ important background, which can be investigated in detail, is the process of exclusive associative charm photoproduction, with pseudoscalar and vector charmed mesons in the final state,  $\gamma+p\to {\cal B}_c+\overline{D}(\overline{D^*})$,
with ${\cal B}_c=\Lambda_c^+$ or $\Sigma_c^+$. The mechanism of photon-gluon fusion, which successfully describes the inclusive spectra of $D$ and $D^*$ mesons at high photon energies, can not be easily applied to exclusive processes, at any energy. Threshold considerations of such processes in terms of standard perturbative QCD have been applied to the energy dependence of the cross section of $\gamma+p\to p+J/\psi$ \cite{Br00}. However, in such approach, polarization phenomena can not be calculated without additional assumptions.

In this respect, the formalism  of the effective Lagrangian approach (ELA) seems very convenient for the calculation of exclusive associative photoproduction of charmed particles, such as $\gamma+p\to \Lambda_c^+(\Sigma_c^+)+\overline{D^0}(\overline{D^*})$ \cite{Re1,Re2,Re3,Re4} at least in the near threshold region. Such approach is also widely used for the analysis of various processes involving charmed particles \cite{Li00}, as, for example, $J/\psi$-suppression in high energy heavy ion collisions, in connection with quark-gluon plasma transition 
\cite{Ma86}. We analyzed earlier different exclusive processes of associative charm particles photoproduction, $\gamma+p\to \Lambda_c^+(\Sigma_c^+)+\overline{D^0}(\overline{D^*})$ and indicated that polarization phenomena can be naturally predicted for those reactions. In this paper we apply the ELA to the collision of circularly polarized photons with a  polarized proton target, starting from the reaction threshold \footnote{Note that beams of high energy circularly polarized photons are actually available at JLab and have been used in the study of helicity conservation in deuteron photodisintegration at high energy \cite{Wi01}}.

%%%%%%%%%%%%%%%%%%%%%%%%%%%%%%%%%%%%%%%%%%%%%%%%%%%%%%%%%%%%%%%%%%%%%%%%%%%%%%%%
\section{The asymmetries for $\overline{D}$ photoproduction on proton}
%%%%%%%%%%%%%%%%%%%%%%%%%%%%%%%%%%%%%%%%%%%%%%%%%%%%%%%%%%%%%
The asymmetry $\cal A$, for the collision of circularly polarized photons with a polarized proton target, $\vec\gamma+\vec p\to \overline{D^0}+X$ is defined as:
\begin{equation}
{\cal A}=\displaystyle\frac{\sigma_{\uparrow\downarrow}-\sigma_{\uparrow\uparrow}}
{\sigma_{\uparrow\downarrow}+\sigma_{\uparrow\uparrow}},
\label{asym}
\end{equation}
where $\sigma_{\uparrow\uparrow}$ $(\sigma_{\uparrow\downarrow})$ is the cross section of $\vec\gamma+\vec p$-collisions with parallel (antiparallel) spins (along the direction of the photon beam, taken as the $z-$direction).

Simple kinematical considerations show that this asymmetry for exclusive photoproduction processes, such as $\vec\gamma+\vec p\to\Lambda_c^+(\Sigma_c^+)+\overline{D^0}$ is sizeable. For example, in the $S$-wave approximation, which applies to the near threshold region, the $\vec\sigma\cdot\vec e$-spin structure of the corresponding amplitudes, (where $\vec e$ is the photon polarization vector) induces a maximal asymmetry, ${\cal A}=+1$, independently on the $\overline{D^0}$ production angle in the reaction CMS. This result is correct for any value of the $S-$wave threshold amplitude and can be directly obtained from the conservation of spin projection in threshold conditions.

In the general case, any reaction $\vec\gamma+\vec p\to
{\cal B}_c+\overline{D^0}(\overline{D^*})$ is described by two different asymmetries and the dependence of the differential cross section on the polarization states of the colliding particles can be parametrized as follows (taking into account the P-invariance of the electromagnetic interaction of charmed particles):
\begin{equation}
\displaystyle\frac{d\sigma}{d\Omega}(\vec\gamma\vec p)=\left ( \displaystyle\frac{d\sigma}{d\Omega}\right)_0
\left (1-\lambda P_x{\cal A}_x-\lambda P_z{\cal A}_z\right ),
\label{sig}
\end{equation}
where $\lambda =\pm 1$ is the photon helicity, $P_x$ and $P_z$ are the possible components of the proton polarization $\vec P$, and ${\cal A}_x$, ${\cal A}_z$ are the two independent asymmetries. Due to the T-even nature of these asymmetries, they are non vanishing in ELA, where the photoproduction amplitudes are real. We will use the standard parametrization \cite{Ch57} of the spin structure for the amplitude of pseudoscalar meson photoproduction on the nucleon:
$${\cal M}(\gamma N\to {\cal B}_c\overline{D})=\chi_2^{\dagger}\left [i\vec\sigma\cdot\vec e f_1 + 
\vec\sigma\cdot\hat{\vec q}\vec\sigma\cdot\hat{\vec k}\times\vec e f_2+ 
i\vec e \cdot\hat{\vec q}\vec\sigma\cdot\hat{\vec k}f_3
+i\vec\sigma\cdot\hat{\vec q} \vec e \cdot\hat{\vec q}f_4\right ]\chi_1,$$
$\hat{\vec k}$ and
$\hat{\vec q}$ are the unit vectors along the three-momentum of $\gamma$ and 
$\overline{D}$; $f_i$, $i$=1-4, are the scalar amplitudes, which are functions of two independent kinematical variables, $s$ and $\cos\theta$, where $\theta$ is the $D-$meson production angle in the reaction CMS, $\chi_1$ and  $\chi_2$ are the two-component spinors of the initial nucleon and the produced ${\cal B}_c$-baryon.

After summing over the ${\cal B}_c$-polarizations, one finds the following expressions for the asymmetries ${\cal A}_x$ and ${\cal A}_z$ in terms of 
the scalar amplitudes $f_i$:
\begin{equation}
{\cal A}_x{\cal D}=    \sin\vartheta {\cal R}e \left [-f_1f_3^*+f_2f_4^*+\cos\vartheta 
(-f_1f_4^*+f_2f_3^*)\right ],
\label{asymx}
\end{equation}
\begin{equation}
{\cal A}_z{\cal D}=   |f_1|^2+|f_2|^2-2\cos\vartheta {\cal R}e f_1f_2^*+ \sin^2\vartheta{\cal R}e \left (f_1f_4^*+f_2f_3^*\right ),
\label{asymy}
\end{equation}
where
\begin{eqnarray}
&{\cal D}=|f_1|^2+|f_2|^2-2\cos\vartheta {\cal R}e f_1f_2^*+ \nonumber\\
&\sin^2\vartheta 
\left \{  \displaystyle\frac{1}{2}\left ( |f_3|^2+|f_4|^2 \right )+  {\cal R}e \left [ f_2f_3^*+2 \left (f_1+\cos\vartheta 
f_3\right ) f_4^*\right ] \right\} .
\label{eq:asymd}
\end{eqnarray}
Note that  ${\cal A}_x$ vanishes at $\vartheta=0^0$ and $\vartheta=\pi$. Moreover ${\cal A}_z=1$, for $\vartheta=0^0$ and $\vartheta=\pi$, for any photon energy. This is a model independent result, which follows from the conservation of helicity in collinear kinematics. It is independent from the dynamics of the process, its physical meaning is that the collision of $\gamma$ and $p$ with parallel spins can not take place for collinear regime. This result holds for any process of pseudoscalar meson photo-production on a nucleon target. But for $\vartheta\ne 0$ and $\vartheta \ne \pi$ the results for ${\cal A}_x$ and ${\cal A}_z$ are model dependent. To estimate the $\cos\vartheta$ and the $E_{\gamma}$-dependences of these asymmetries we use the ELA model. The explicit expressions for the amplitudes $f_i,i=1-4$, in terms of the $g_{N{\cal B}_cD}$ coupling constant and the magnetic moments of the charmed baryons, in framework of ELA, can be found in \cite{Re1}.

The starting point of this model is the $c$-quark exchange (through the $u$-channel ${\cal B}_c$-exchange). Other diagrams (D-exchange in $t-$channel and $N$-exchange in $s$-channel) have to be included because the matrix element of this process has to satisfy the gauge invariance. It is this symmetry of the electromagnetic interaction which drives the relative role of different possible mechanisms in $\gamma+N\to{\cal B}_c+\overline{D}$, and therefore the angular distributions and the $\Sigma_{\cal B}$-asymmetry induced by linearly polarized photons colliding with unpolarized target. In the near threshold region, the main contribution is represented by the $u$-channel mechanism \cite{Re2}. This mechanism occurs for charm photoproduction at short distances, $\simeq 1/m_c$, where $m_c$ is the $c$-quark mass, in agreement with a previous study of threshold exclusive $J/\psi$ photoproduction \cite{Br00}. This is the typical radius for exclusive charmed particle photoproduction, and a larger radius (up to 1 fm) can be achieved only for more complicated processes, where the light mesons exchanges are possible, such as $\gamma+p\to D+\overline{D}+N$.

Applying this model in \cite{Re2}, we found  the angular and the energy dependence of the differential cross section for the processes 
$\gamma+N\to{\cal B}_c+\overline{D}$ and compared the results with the existing experimental data on open charm photoproduction. The absolute values of the corresponding cross sections are determined by the $g_{N{\cal B}_cD}$ coupling constants, which are presently not known, and have been considered as adjustable parameters. The deduced values for these coupling constants are too low in comparison with the predictions of $SU_4$-symmetry. Such conclusion has been confirmed by our analysis  \cite{Re4} of vector meson $D^*-$photoproduction.

As  the constant $g_{N{\cal B}_cD}$ can not be independently determined, the relative contribution of individual channels to the inclusive open charm photoproduction can not be assessed. However, a crude estimation of this contribution can be done from the measured charm-anticharm asymmetry in open charm photoproduction, because the exclusive processes generate $\overline{D}$-mesons and charmed baryons. From the observed $\Lambda_c/\overline{\Lambda_c}$-asymmetry \cite{Bi99}, one can estimate a contribution of exclusive photoproduction of open charm of the order of 20\%, which may seriously affect the experimental $D-$asymmetry in polarized $\vec\gamma\vec p$-collisions.

\begin{figure}
\mbox{\epsfysize=15.cm\leavevmode \epsffile{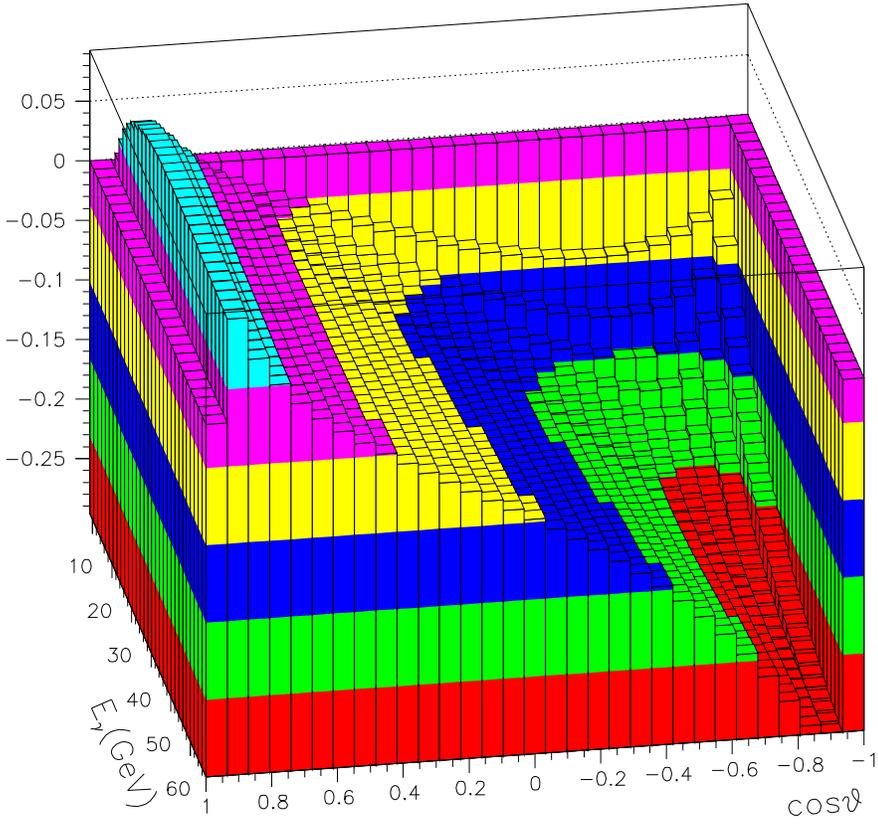}}
\vspace*{.2 truecm}
\caption{$\cos\vartheta$- and $E_\gamma$ dependencies of the asymmetry 
${\cal A}_z$, for $\vec\gamma+\vec p\to
{\Lambda}_c^++\overline{D^0}$, calculated for $\mu_{\Lambda_c}$=0.37 $\mu_N$
\protect\cite{Sa94}.}
\label{fig:fig1}
\end{figure}

\begin{figure}
\mbox{\epsfysize=15.cm\leavevmode \epsffile{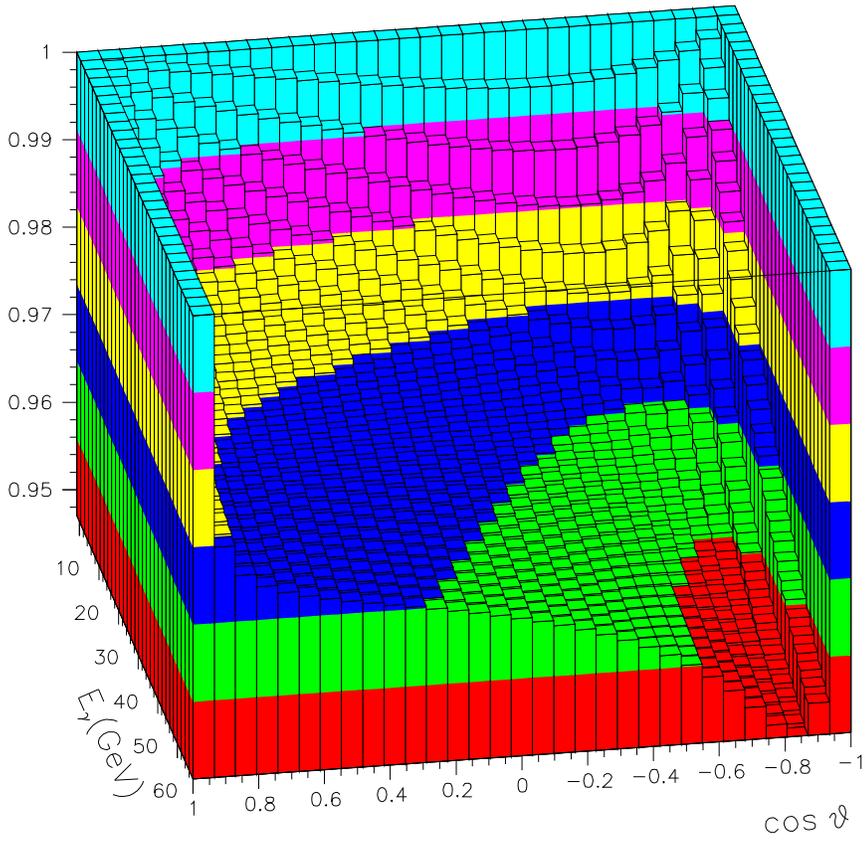}}
\vspace*{.2 truecm}
\caption{Same as Fig. 1, for the asymmetry  ${\cal A}_x$.}
\label{fig:fig2}
\end{figure}

\begin{figure}
\mbox{\epsfysize=15.cm\leavevmode \epsffile{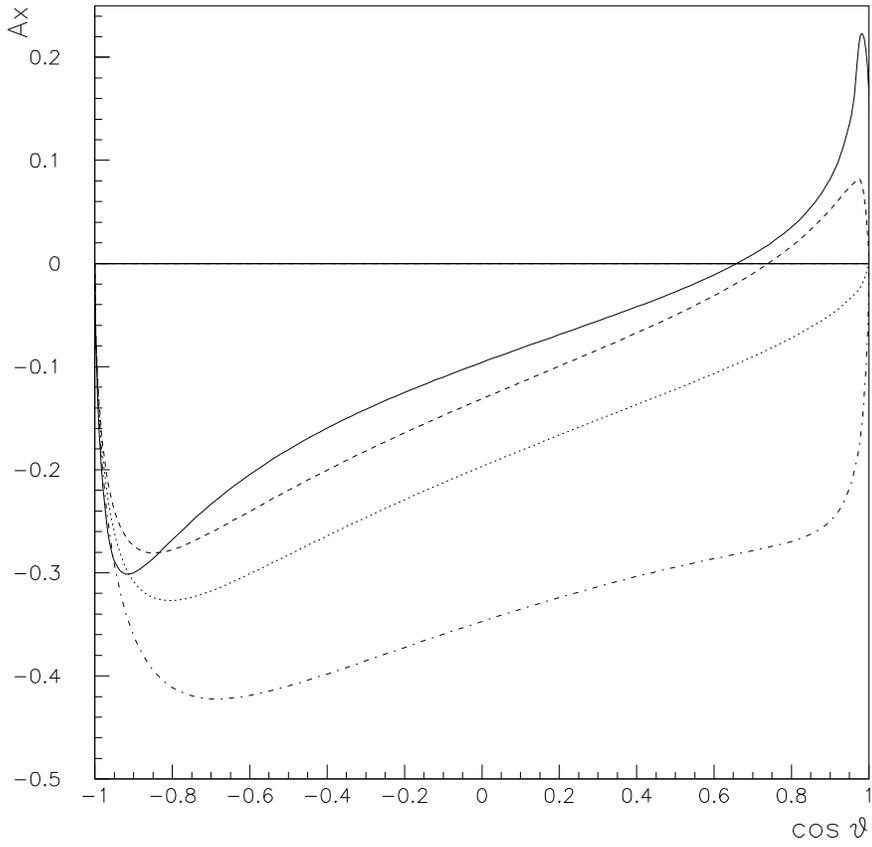}}
\vspace*{.2 truecm}
\caption{$\cos\vartheta$-dependence of the asymmetry ${\cal A}_x$, at 
$E_\gamma$=40 GeV, for different values of $\mu_{\Lambda_c}$ (in units of $\mu_N$): $\mu_{\Lambda_c}$=-1 (solid line), $\mu_{\Lambda_c}$=0.5 (dashed line), $\mu_{\Lambda_c}$=1 (dotted line), and $\mu_{\Lambda_c}$=2 (dash-dotted line)}
\label{fig:fig3}
\end{figure}

The asymmetries ${\cal A}_x$ and ${\cal A}_z$ do not depend on the coupling constant $g_{N{\cal B}_cD}$, in the framework of the considered model, but they  depend on $\kappa_{{\cal B}_c}$, the anomalous magnetic moment of the charmed baryons. The energy and the angular dependences of ${\cal A}_x$ and ${\cal A}_z$, for the process $\vec\gamma+\vec p\to
{\Lambda}_c^+ +\overline{D^0}$ are shown in Figs. \ref{fig:fig1} and \ref{fig:fig2}.  The asymmetry ${\cal A}_z$ is large, ${\cal A}_z\ge 0.9$, in all the considered kinematical region with small sensitivity to the 
$\Lambda_c^+$-magnetic moment. The asymmetry ${\cal A}_x$ is negative in most part of the considered kinematical region and smaller in absolute value. Note also that the asymmetry ${\cal A}_x$ is more sensitive to the $\Lambda_c^+$-magnetic moment (see Fig. \ref{fig:fig3}).

\section{Photoproduction of vector mesons}

The vector meson $D^*$-photoproduction can be considered in a similar way. As in case of photoproduction of pseudoscalar mesons,  the asymmetry for polarized $\vec\gamma+\vec p$- collisions takes the maximal value near threshold. Let us write the most general parameterization of the spin structure of the amplitudes for $\gamma+N\to {\cal B}_c+\overline{D^*}$, with $S$-wave production of final particles:

\begin{eqnarray}
&{\cal M}(\gamma N\to &{\cal B}_c \overline{D^*})= \chi_2^{\dagger}
\left [ e_1 \left (\vec e\cdot\vec U^*+i\vec\sigma\cdot\vec U^*\times\vec e\right )+ 
\right . \nonumber \\
&&\left . e_3\left (2\vec e\cdot\vec U^*-i\vec\sigma\cdot\vec U^*\times\vec e\right )
+im_3\left (\vec\sigma\cdot\vec e\times\hat{\vec k}\vec U^*\cdot\hat{\vec k}+
 \vec\sigma\cdot\hat{\vec k}\vec e\times\hat{\vec k}\cdot\vec U^*\right )\right ]\chi_1,
\label{mdst}
\end{eqnarray}
where $\vec U$ is the three-vector of $\overline{D^*}$-polarization, $e_1$ and $e_3$ are the multipole amplitudes corresponding to the absorption of electric dipole photons with production of final particles in states with spin ${\cal J}$ and parity $P$, ${\cal J}^P=1/2^-$ and $3/2^-$ respectively, and $m_3$ is the multipole amplitude for absorption of magnetic quadrupole photon, so ${\cal J}^P=3/2^-$.

Only the asymmetry ${\cal A}_z$ does not vanish, at the reaction threshold, and it can be written as a function of the threshold amplitudes $e_1$, $e_3$ and $m_3$:
\begin{equation}
{\cal A}_z(\gamma p\to \Lambda_c^+\overline{D^*})=  \displaystyle\frac{3|e_1|^2-3|e_3|^2+|m_3|^2-2 {\cal R}e (2e_1+e_3^*)m_3^*}
{3|e_1|^2+6|e_3|^2+2|m_3|^2}.
\label{asymzd}
\end{equation}
For pure multipole transitions, with $e_1$, $e_3$ or $m_3$ amplitude, we have, respectively 
${\cal A}_z(\gamma p\to \Lambda_cD^*)= 1$, -1/2 and 1/2.

A model for $D^*$-photoproduction, based on the pseudoscalar $D$-exchange, has been considered in \cite{Re1}. Such mechanism gives ${\cal A}_x=0$ and ${\cal A}_z$=0, for any photon energy and any production angle, because no spin correlation between the polarizations of $\gamma$ and $p$ can be induced by the exchange of a zero spin particle. Therefore the discussed asymmetries can be considered a powerful signature of the possible exchange mechanism. Note that these results do not depend on the exact values of the corresponding coupling constants and the form of the phenomenological hadronic form factors, which are typical ingredients of such calculations.

On the basis of our previous consideration \cite{Re4} of possible baryon exchange contribution to the threshold multipole amplitudes $e_1$, $e_3$ or $m_3$, one can derive the following formula for the asymmetry  
${\cal A}_z(\gamma p\to \Lambda_c^+\overline{D^*})$:
\begin{equation}
N{\cal A}_z(\gamma p\to \Lambda_c^+\overline{D^*})=
\displaystyle\frac{r}{M}
\left \{ -2(W-m)+
\displaystyle\frac{r}{M}\left [ (M+m)^2-2m_{D^*}
\left (W+M \right )\right ]\right \},
\label{eq:asym1}
\end{equation}
with
$$N=\left (\displaystyle\frac{W-m}{m_{D^*}}\right )^2+2r\displaystyle\frac{(W-m)(W+M)}{m_{D^*}M}+\left (\displaystyle\frac{r}{M}\right )^2\left (
m^2+2W^2+2mW+3M^2 \right ),$$
where $W=M+m_{D^*}$ and $m$, $M$, $m_{D^*}$ are the masses of nucleon, $\Lambda_c$-hyperon and $D^*$-meson, respectively. The real parameter $r$ characterizes the ratio of the coupling constants \footnote{The parameter $r$ determines the relative role of $D$-exchange mechanism and baryon-exchange mechanisms for $\gamma+p\to\Lambda_c^+ +\overline{D^{*0}}$ \protect\cite{Re4}.}:
$$r=2\displaystyle\frac{g_v}{g_{D^{*0}D^0\gamma} g_{p\Lambda_cD} }.$$
where $g_v$ is the vector coupling constant for the vertex 
$p\to \Lambda_c^+ +\overline{D^{*0}}$ and $g_{D^{*0}D^*\gamma}$ is the magnetic moment for the $D^{*0}\to D^{0}$ transition. 
This moment determines the width of the radiative decay $D^{*0}\to D^{0}+\gamma$, with experimental branching ratio $\simeq 40\%$. Only the upper limit is experimentally known for the total width of the neutral $D^{*0}$:
$\Gamma (D^{*0})\le $2.1 MeV \cite{PdG}.  The total width 
of the charged $D^{*+}$-meson is determined from the recent measurements of the CLEO collaboration \cite{PdG}: $\Gamma (D^{*+})= (9.6\pm 4\pm 2.2)$ keV.  Using the known value for the branching ratio of the radiative decay $\Gamma(D^{*+})\to D^* +\gamma$ \cite{CLEO2}:
$$BR=(16.8\pm 0.42\pm 0.49 \pm 0.03)\%,$$
one can find a value of the magnetic moment for the transition $D^{*+}\to D^+ +\gamma$, which agrees with the predictions from different theoretical approaches. Therefore it seems reasonable to use these models for the calculation of the coupling constant $g_{D^{*0}D^0\gamma}$. But the main problem, in our case, is how to determine the coupling constants for the vertices $N\Lambda_cD$ and $N\Lambda_cD^*$. The $SU(4)$-symmetry, which connects these coupling constants with the corresponding coupling constants for the vertices $N\Lambda K$ and 
 $N\Lambda K^*$  is essentially violated \cite{Re3}. We showed earlier \cite{Re4}, that the parameter $r$ can be determined, in principle, in the reaction $\gamma +p\to \overline{D^{*0}}+\Lambda_c^+$ through the angular dependence of the pseudoscalar $D$-meson produced in the subsequent decay $D^{*0}\to D+\pi$. 
 
\begin{figure}
\mbox{\epsfysize=15.cm\leavevmode \epsffile{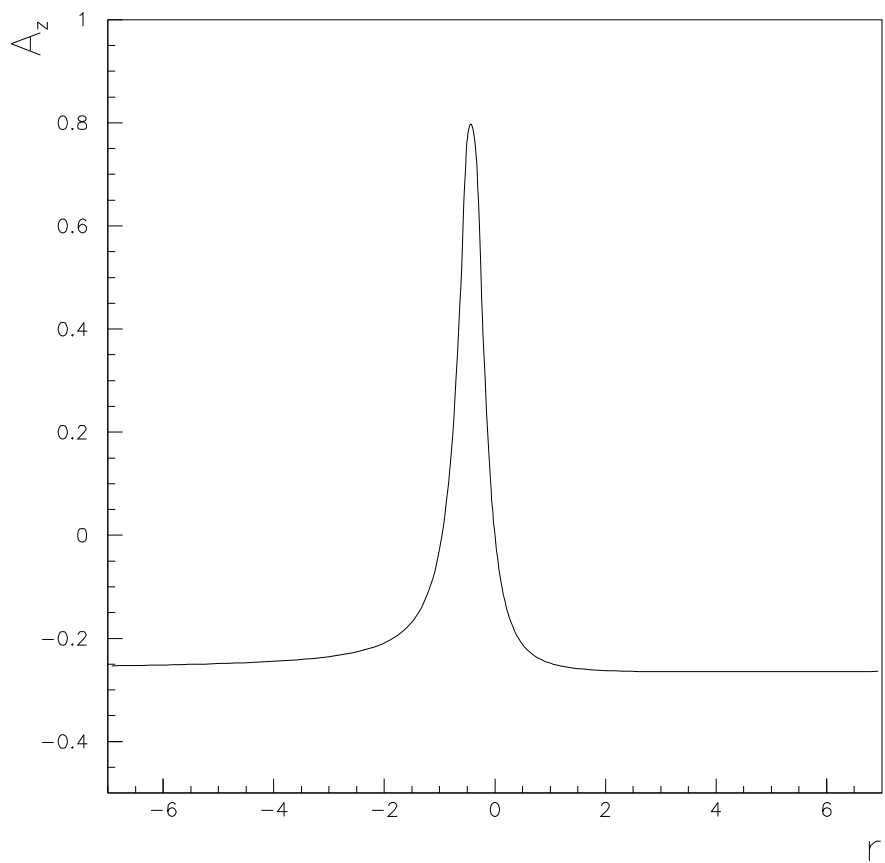}}
\vspace*{.2 truecm}
\caption{$r$-dependence of the threshold asymmetry, ${\cal A}_z$ ($\vec\gamma\vec p\to \Lambda_c D^*$) }
\label{fig:fig4}
\end{figure}

Fig. \ref{fig:fig4} shows the dependence of the asymmetry ${\cal A}_z$, Eq. (\ref{eq:asym1}), on the ratio $r$. One can see that only for $r\simeq -1$ the asymmetry ${\cal A}_z(\vec\gamma \vec p\to \Lambda_c^+\overline{D^*})$  is positive, being generally negative, ${\cal A}_z\simeq -0.25$, with very small sensitivity to the ratio $r$ of the fundamental coupling constants.

From this brief discussion it appears that the prediction of the asymmetry ${\cal A}_z$ for charmed vector meson photoproduction is more complicated and more model dependent in comparison with processes involving  pseudoscalar mesons, as this asymmetry depends on the value of $r$. We can suggest the study of the asymmetry in $\vec \gamma+\vec p\to \Lambda_c^++\overline{D^*}$ as a possible way to determine the fundamental coupling constants for the vertices involving charmed particles.

The model of photon-gluon fusion predicts a value of $\simeq 15\div 30 \%$, for the ${\cal A}_z$-asymmetry in the inclusive $\vec\gamma +\vec p\to \overline{D^{0}}+X$ process, at $E_\gamma\simeq$ 50 GeV, depending on the assumptions on the polarized gluon distribution, $\Delta G$ \cite{Compass}. Therefore, even a $10\%$ contribution of the exclusive process $\vec \gamma+\vec p\to \Lambda_c^++\overline{D^0}$ to the inclusive $D$-cross section can induce a $10\%$ correction to the $\gamma G$-fusion asymmetry, at forward angles. It is a very large effect, which should be taken into account in the extraction of the $\Delta G$-effect from the asymmetry in the process $\vec \gamma+\vec p\to \overline{D^0}+X$. Evidently this effect does not contribute to the  $D^0$ production in the process $\vec \gamma+\vec p\to D^0+X$.

The observation of the $\overline{D^0}/D^0$ or $\overline{\Lambda_c}/\Lambda_c$ asymmetry in $\gamma N$-collisions will be important in order to test the validity of the photon-gluon fusion mechanism.

Note that the exclusive photoproduction of open charm will result in $\overline{D^0}/D^0$ and $\overline{\Lambda_c}/\Lambda_c$ asymmetry in $\gamma N$-collisions (with unpolarized particles). Such asymmetries have been experimentally observed. For example, the FOCUS experiment found a ratio for $\overline{\Lambda_c}/\Lambda_c$ production $\simeq 0.14\pm 0.02$, demonstrating that $\Lambda_c^+$-production is more probable than $\overline{\Lambda_c}$ \cite{Bi99}.  This is in contradiction with the model of 
$\gamma G$-fusion, which predicts symmetric $\overline{\Lambda_c}-\Lambda_c$-yields and can be considered as an indication of the presence of other mechanisms. The exclusive processes, discussed in this paper, explain naturally such asymmetry. Similar charm asymmetries have been observed also at $E_{\gamma}$=20 GeV, at SLAC \cite{Ab84}.

\section{Conclusions}

We calculate the T- and P-even asymmetries for the processes $\vec\gamma+\vec p\to \Lambda_c^+ (\Sigma_c^+)+\overline{D^0}$ and $\vec\gamma+\vec p\to \Lambda_c^+ (\Sigma_c^+)+\overline{D^*}$, induced by 
circularly polarized photons on a polarized proton target. Using an effective Lagrangian approach, we found that these asymmetries are large in absolute value for all considered photon energies. For the pseudoscalar meson production the ${\cal A}_z$ asymmetry is essentially positive,  large in magnitude and of the same sign of the predictions based on the photon-gluon fusion mechanism. This means that exclusive processes have to be viewed as a non negligeable contribution to charm photoproduction at high energies and they have to be carefully treated in order to extract the polarization properties of the gluon content of the nucleon from the future Compass data.

For vector charmed meson photoproduction the predictions for the discussed asymmetries are more model dependent. Vanishing for the $D-$meson exchange mechanism, the threshold asymmetry ${\cal A}_z$ shows large sensitivity to the ratio of the coupling constants for the different vertices involving charmed particles.

%\section{Aknowledgments}

%%%%%%%%%%%%%%%%%
%%%Figure 1 %%%%%
%%%%%%%%%%%%%%%%%

\end{document}